\documentclass[12pt,preprint]{aastex}
\shorttitle{A Galactic Stellar Structure in Triangulum-Andromeda}
\shortauthors{Majewski et al.}
\usepackage{psfig}

\begin{document}

\title{Detection of the Main Sequence Turn-off of a Newly Discovered
Milky Way Halo Structure in the Triangulum-Andromeda Region}

\author{Steven R.  Majewski\altaffilmark{1}, 
James C. Ostheimer\altaffilmark{1,2},
Helio J. Rocha-Pinto\altaffilmark{1},
Richard J. Patterson\altaffilmark{1},
Puragra Guhathakurta\altaffilmark{3},
David Reitzel\altaffilmark{4}  }

\email{srm4n@virginia.edu, jostheim@alumni.virginia.edu, hjr8q@virginia.edu, rjp0i@virginia.edu,
  raja@ucolick.org, reitzel@astro.ucla.edu}

\altaffiltext{1}{Astronomy Dept., Univ.\ of Virginia, Charlottesville,
  VA 22903-0818}

\altaffiltext{2}{Current address: 1810 Kalorama Rd. NW, \#A3, Washington DC 20009}

\altaffiltext{3}{UCO/Lick Observatory, Dept. of Astronomy \& Astrophysics, Univ. of California, 
Santa Cruz, CA 95064}

\altaffiltext{4}{Dept. of Physics and Astronomy, Univ. of California, Los Angeles, CA 90095}

\begin{abstract}

An upper main sequence (MS) and 
main-sequence turn-off (MSTO) feature appears in the color-magnitude diagram (CMD)
of a large area photometric survey of the southern half of M31 stretching to M33.  
Imaging in the Washington $M,T_2,DDO51$ system allows
us to remove the background M31/M33 giants from our CMD and more clearly
defined the dwarf star feature, which has an MSTO near $M\sim20.5$.  The corresponding
stellar population shows little density variation over the $12^{\circ} \times 6^{\circ}$
area of the sky sampled and
is of very low surface brightness, $\Sigma > 32$ mag arcsec$^{-2}$.
We show that this feature is not the same as a previously identified, 
MS+MSTO in the foreground of the Andromeda Galaxy that has 
been associated with the tidal stream ringing the Milky Way disk at less than
half the distance. 
Thus, the new stellar system is a separate, more distant entity,
perhaps a segment of tidal debris from a disrupted satellite galaxy.
It is most likely related to the structure with similar distance, location
and density uniformity seen as an excess of K and M giants in the
Two Micron All-Sky Survey reported in the companion paper by Rocha-Pinto et al. (2004).
 

\end{abstract}
\keywords{Galaxy: structure --- Galaxy: halo ---  Galaxy: evolution ---
galaxies: interactions --- galaxies: Local Group --- galaxies: dwarf}

\section{Introduction}

The search for halo substructure has become an important avenue by which to
constrain models for the growth of galaxies and their stellar populations, which is
currently thought to occur through the accretion of smaller companions (e.g., Searle \& Zinn
1978, White \& Rees 1978).  Because this substructure is typically subtle and obscured by a 
substantial foreground veil of disk stars, eliciting its presence requires strategies that
optimize the substructure signal compared to the foreground noise.  These strategies include
searching for stellar tracers particular to the halo substructure at certain apparent
magnitudes, like RR Lyrae stars (e.g., Vivas et al.\ 2001) or M giants 
(e.g., Majewski et al.\ 2003), or utilizing specific regions of the color-magnitude diagram
where the presence of the substructure is expected to be strongly expressed compared
to the background (e.g., Kuhn, Smith \& Hawley\ 1996, Rockosi et al.\ 2002).  The 
main sequence-turnoff (MSTO) region has proven particularly useful for identifying or tracing out
the tidal tails of accreted satellite systems in the Galactic halo
(e.g., Grillmair et al.\ 1995; Mart{\'{\i}}nez-Delgado et al.\ 2001, 2004; Newberg et al.\ 2002, 2003;
Ibata et al.\ 2003, hereafter ``I03").  The latter five references have each inferred the presence of
new tidal debris structures on the basis of identified MSTO structures in field color-magnitude
diagrams.  The present discussion follows the precedent 
of these other papers. 

In the course of his deep, large area survey of M31 giants, Ostheimer
(2002; see also Ostheimer et al.\ 2004) noted a 
feature in his integrated color-magnitude diagram (CMD) that resembled a
main sequence turn-off for a foreground Milky Way population.  The goal of this
paper is to report that this MSTO must be related to a new halo substructure
in the direction of Tringulum-Andromeda.  I03 have 
previously reported the existence of an MSTO feature in the foreground of
the Andromeda galaxy that they associate with the ring-like tidal stream 
around the Galactic disk discovered by Newberg et al.\ (2002)
and described by Yanny et al.\ (2003) and others.  We show that  
the new stellar population reported here is about two magnitudes
fainter and at least twice as distant as the ring feature previously described.


\section{Data}

A full description of the data and their reduction is contained in Ostheimer
(2002), Ostheimer et al.\ (2004), and Ostheimer et al.\ (in preparation).
Ten ``M31 halo" fields stretching from the southeast of M31 to the northwest of M33 were 
surveyed using the MOSAIC camera on the KPNO 4-m telescope on the nights of 
UT 14-18 Nov 1998, and UT 26 Dec 2001-02 Jan 2002.
The field placement is shown
in Figure 1.  Photometry was obtained in the Washington $M$, $T_2$ and $DDO51$ filters, 
a combination employed in the Ostheimer et al.\ studies 
to separate foreground Milky Way dwarf stars from M31 halo giants
using the technique described by Majewski et al.\ (2000; see below).  
Stellar sources were identified and photometered using DAOPHOT II (Stetson 1992) and 
ALLFRAME (Stetson 1994); Figure 2 (bottom panel)
shows the combined color-magnitude diagram (CMD) for more than 8.3 million stellar 
sources (selected with DAOPHOT $\chi \le 1.3$ and $sharp <0.3$ to remove 
galaxies) corrected for
reddening on a star-by-star basis using the maps of Schlegel et al.\ (1998).  The total area represented
is 3.6 deg$^2$.  The large 
mass of points at faint magnitudes in the CMD corresponds to the top of the M31/M33 giant
branch (and perhaps some contribution of contaminating compact galaxies), 
while the smaller mass of points redder than $(M-T_2)_0 > 2.5$ is from nearby
late type dwarf stars.  Here we call attention to the hook-like feature near $(M-T_2)_0 \sim 1$
above the M31 giants, brighter than $M-A_M \sim 23$.  It clearly is not associated
with M31, as we shall now show.  

Because the $M-DDO51$ color is primarily sensitive to stellar surface gravity whereas $M-T_2$ is
primarily sensitive to surface temperature, 
the $(M-T_2, M-DDO51)_0$ diagram (2CD; top panel of Figure 2) has previously been used to remove foreground
dwarf stars as contaminants from distant giant star samples (e.g., Majewski et al.\ 2000b, Palma
et al.\ 2002); but here the foreground
dwarfs are the targets of interest.  To remove the background giant stars, we fit the dwarf star 
locus in the 2CD (see Majewski et al.\ 2000) of Figure 2 
with an eighth order polynomial and discard all 
stars with an $M-DDO51$ color more than 0.075 mag from that locus.  We also impose
magnitude error limits of 0.075 in all passbands for the present paper.  A CMD 
of the resulting sample of ``dwarf only" stars (Fig.\ 3a) shows the hook-like feature more
clearly, and 
largely free of the background M31 giant stars and compact galaxy contamination.\footnote{Galaxies 
not already removed by the DAOPHOT morphological parameters 
are generally removed by the 2CD analysis because galaxy spectral energy distributions
have a strong contribution 
from K or M giant stars, and, moreover, galaxies at larger redshifts have their Mgb+MgH feature
redshifted out of the $DDO51$ passband, giving them an $M-DDO51$ color identical to that
of a featureless object at $DDO51$ wavelengths.}
That the MS feature appears
to ``end" at $M_0 \sim 23$ is an artifact of the irregular limiting magnitudes and seeing
among the ten M31 fields.  

To assess the significance of the MS feature we must first account for the
``background'' of other, unrelated sources.  Unfortunately, because the MS+MSTO feature
appears in all of our fields, we have no CMD to use as a ``control field" to
subtract from Figure 3a.  Thus, an estimate the number of ``background" stars falling
in the region of the MS must be estimated from the density of stars in nearby regions of the
CMD.  We prefer to avoid regions of the CMD where there is a less reliable
handle on the ``CMD background" level, and this forces us away from using the MSTO proper.
Instead, focus is placed on the region in the Figure 3a CMD where the MS feature is both
distinct and for which there are populated regions of the CMD to either side of the MS
from which to interpolate, namely $21.5 \le M_0 \le 23.0$.
As a tool, we include in Figure 3b a ridge line ($M_0 = 5.0 [M-T_2]_0 + 17.5$) 
for the MS in this magnitude range.  This particular ridge line was found to give
a minimum width in the distribution of differences between the colors of stars in 
the MS structure and the ridge line over $21.5 \le M_0 \le 23.0$ (Fig.\ 4a); this ridge
line is 
purely observationally derived and for now we ascribe no special astrophysical content to it. 
Figure 4 shows the distribution of $(M-T_2)_0$ color differences from the ridge line
for $21.5 \le M_0 \le 23.0$.  The peak near the origin of Figure 4a represents the 
MS in this magnitude range.  To estimate the background in the region of the MS, we 
fit a fourth order function to the two 0.30 color ranges to either side of, and not part of, the peak.
The inset to Figure 4a shows the resulting distribution within the region defining the
background.  By this analysis, over the ten fields in our survey there are 934 stars
in the $-0.30$ to $+0.25$ color difference range, while the background is estimated to be
335.9 stars.  The result is that {\it over this 1.5 magnitude range of the MS} there
are estimated to be $598.1\pm30.6$ MS stars (with the error derived from Poissonian statistics),
and the MS is detected with a $S/N \sim 20$ {\it at these magnitudes}.

\section{Exploring The Main Sequence Turn-Off Feature}

The thickness of the MSTO feature in Figures 2 and 3, and the flat-topped distribution
in Figure 4a, are deceptive --- the product
of the large span of sky surveyed.  In fact, the MSTO feature is thinner in any one of our
0.36 deg$^2$ fields, but appears to change distance modulus across the fields, generally getting 
farther with higher $|b|$.  CMDs of more limited spans of sky area (Figure 5) 
demonstrate thinner, more coherent MSTOs 
as well as the distance modulus shift as a function of Galactic latitude --- in this case
between fields m1, a0 and m4 spanning
Galactic latitudes $-22^{\circ}>b>-24^{\circ}$ (Fig.\ 5a) compared to the fields
m11, b15, and a19 spanning 
$-27^{\circ}>b>-28.1^{\circ}$ (Fig.\ 5b).  The shift is also evident in Figures 4b and 4c
where the color difference distribution is compared for the two groups of fields shown in Figure 5a and 5b, 
respectively: The median color difference between the two samples is 0.075 dex,\footnote{This is 
more than $5\times$ the uncertainty in the difference, 
where the uncertainties in the median values are estimated 
by $\sigma/N^{0.5}$ with $\sigma \sim 0.13$ conservatively estimated by the best Gaussian
fit to the broader distributions, {\it ignoring} the narrow peaks in each.} and, 
adopting the form of the Figure 3 ridge line, this corresponds to a difference in distance modulus
between the two Galactic latitudes of $0.37\pm0.07$ magnitudes (a difference in distance of 19\%), assuming
no metallicity differences. 
Figures 4 and 5 show that the MS feature has more coherence than suggested in Figures 2 and 3, and that
the width of the feature in those latter figures is due neither to a substantial metallicity spread
nor a line-of-sight spread in the stellar population creating it.  

I03 have previously reported the presence of a foreground
MS+MSTO feature in a number of fields in the sky, including, in particular, fields 
around M31 and M33 (the latter are marked in Fig.\ 1).  However, their focus is on a 
populations having an MSTO feature at brighter magnitudes ($V_0 \sim 18.5$),
which they attribute to the
ring-like tidal stream first discovered in Monoceros by Newberg et al.\ (2002;
see also Yanny et al.\ 2003) and found to extend to greater angles around
the sky by I03 as well as by Majewski et al.\ (2003), Rocha-Pinto et al.\ (2003), Crane
et al.\ (2003), Martin et al.\ (2003) and Frinchaboy et al.\ (2004).  
We convert the ridge line of their discovered MSTO feature
into the Washington system\footnote{Majewski et al.\ (2000) give the
transformation between Washington $M$ and $M-T_2$ and Johnson-Cousins $V$ and
$V-I_C$.  The I03 CMDs of M31 are in the Johnson $V$
and Gunn $i'$ band of the Sloan Digital Sky Survey.  From the
Wide Field Survey website, http://www.ast.cam.ac.uk/$\sim$wfcsur/colours.php,
we obtain the conversion from $(i', V-i')$ to the Landolt
$(V, V-I)_L$ system.  Menzies et al.\ (1991) point out a slight difference between
the Landolt and the true Cousins $(V, V-I)_C$ systems;   
we account for this difference before translating to the
Washington system.  Thus, we obtain $(M-T_2) = 0.007 + 1.379(V-i')$ and
$M = V + 0.006 + 0.2(M-T_2)$. }
and include
that ridge line in Figures 3 and 5.  As may be seen, this feature fails to
match the MSTO feature in our data by some 1.5 to 2.5 magnitudes.  

We can rule out that our MSTO feature is not simply a population at the same distance
as the I03 feature, but somehow more subluminous by $1.5-2.5$ magnitudes. 
Girardi et al.\ (2002) have produced isochrones in the Washington system
for populations with a variety of metallicities and ages.  Because our
data are saturated for $M<19$, we do not have 
a clear view of a subgiant/giant branch that would be associated 
with our MS+MSTO feature; access to this would be helpful for determining an
appropriate metallicity/age isochrone to adopt for main sequence fitting.  
However, because metallicity is the predominant parameter driving MS luminosity at the colors
of our ridge line in Figure 3, it is sufficient to test whether a reasonable metallicity
spread can explain the difference between the I03 and our own ridge lines.
Figure 3c shows a spread of isochrones from solar to [Fe/H]$=-2.3$, all shifted
by a distance modulus of 16.0 magnitudes.  We have selected younger isochrones for the 
metal-rich isochrones 
so that they extend to the blue colors of the MSTOs of both the I03 and the newly
identified MS+MSTO features, but this is of little consequence to the
following discussion.  As may be seen, more than this 2.3 dex span of metallicity would be  
needed to place both the I03 and our fainter ridge line at the same distance, and the situation is exacerbated 
in the case of the even fainter MS seqences in some of the Ostheimer fields (e.g., those shown
in Figure 5b).  The argument against association of the two MS populations is
made even stronger by the fact that Yanny et al.\ (2003) have associated the 
ring MS+MSTO feature with a population dominated by [Fe/H]$=-1.6$ stars; if this is the case, it
is not possible that our feature is at the same distance because the 
main sequence of a [Fe/H]$=-1.6$ population is already nearly to the fullest possible
subdwarf displacement relative to solar metallicity stars (as shown by the two bottom isochrones in Fig.\ 3c). 
Clearly the MS+MSTO we have identified belongs to a coherent stellar structure different
than and beyond the foreground system associated with the I03 MS+MSTO. 

One may wonder why there is no strong indication of the I03 ring
feature within our CMDs.  This is due to the strong exponential gradient (described
in terms of a scaleheight with $b$ in I03) in
the density of the ring stars with Galactic latitude: As shown in 
Figure 10 of I03, the density contrast in ring stars
from their northern ($b=-19^{\circ}$) and southern ($b=-24^{\circ}$) 
M31 fields (the latter corresponding to the lowest $|b|$ of our fields)
is about a factor of three to one.
A comparison of their north
and south M31 CMDs (their Figures 6 and 7) clearly shows how diffuse
the ring MS+MSTO feature has become in the CMD by $b=-24^{\circ}$; by their 
M33 field ($b=-31^{\circ}$) I03 detect no ring stars.  
On the other hand, the structural feature we identify shows a much weaker 
density gradient across our fields (e.g., compare the relative
density of the MS+MSTO in Figs. 5a and 5b; see also \S5) {\it and} it is 
plainly visible 
in the M31 field CMDs shown in I03,
though these authors do not comment on this more distant feature appearing in their CMDs.
The existence of {\it two} MS+MSTO features in the I03 M31 CMDs, the fainter
one corresponding to the one we have identified and
roughly 1.5-2.5 magnitudes fainter than the ($\sim8$ kpc distant) MS+MSTO they discuss,
is additional proof that there are two, distinct Milky Way halo structures in this part of the sky. 
A comparison of Figures 6 and 7 of I03 shows an interesting change in density 
constrast between the near and
far MS+MSTO structures at two different Galactic latitudes ($b=-19^{\circ}$ and $-24^{\circ}$):
The two MS+MSTOs are more comparable in density in their $b=-24^{\circ}$ field, but the 
nearer population is much more populous in the $b=-19^{\circ}$ field, for reasons discussed 
above.
 
The Ostheimer and I03 studies  
may not be the only surveys to have uncovered this newly-identified Galactic
star system, although, to our knowledge, it has never been pointed out before.
For example, the MS+MSTO hook we observe here also clearly lies in the ``M31" CMDs
presented by Durrell et al.\ (2001) --- whose fields correspond well
to the location of fields in our study 
(Fig.\ 1) --- but these authors do not bring attention to this MS+MSTO
feature in their CMDs.  More recently Durell et al.\ (2004) have expanded their
survey area, and obtain similar results in terms of a clear signal of the fainter MS+MSTO
hook in their CMDs.
However, the similar {\it lack} of the I03 ring MS+MSTO feature
in the southerly fields of Durrell et al.\ (2004), especially compared to their more 
northern $\mathcal {R}1$ field where a double MS+MSTO is seen (as with the northern I03
field), is consistent with our own
non-detection of the brighter ring MS+MSTO feature in the same general area to the south of
M31.


\section{Discussion}

As discussed above, without a clearly defined subgiant branch it is 
difficult to assign a specific age and metallicity to our MS+MSTO feature, although
a variety of combinations of age and $Z$ can be ruled out (e.g., old, relatively
metal-rich -- say 10 Gyr, $Z>0.001$ --- isochrones do not go blue enough).
What we {\it can} say is that the MS for any specific $Z$ isochrone must
be at a specific distance modulus to match the observed sequence.  For example, as shown in Figure 3c
we find that $Z \le 0.001$ ([Fe/H] $< \sim -1.28$) isochrones must 
be shifted to a distance modulus of 16.0 to match the mean ridge line 
across our fields, whereas a $Z=0.004$ ([Fe/H]$=-0.68$) isochrone
must be shifted to 16.5 and a solar metallicity ($Z=0.019$) isochrone 
must be shifted to 17.0 to match the ridge line.  Thus, with no information
on metallicity, the best we can say about the
distance of our new stellar structure is that it is from 
$\sim16$ to $\sim25$ kpc
away if it is of poor to solar metallicity, respectively.  On top of this,
we identified above (\S3; Figs.\ 4 and 5) an approximately 19\% {\it distance variation}
from field to field, with a larger distance found at higher Galactic latitude.

On the other hand
over the roughly $12^{\circ} \times 6^{\circ} $
area surveyed (Fig.\ 1) the {\it density} of the feature is roughly constant.  
For this comparison we combine fields at similar Galactic latitudes to
improve statistics.
For example, the number of stars in the MS feature from $21.5 \le M_0 \le 23.0$, 
as determined by the background subtraction analysis demonstrated in Figure 4, 
is $183.2\pm16.9$ and $194.3\pm17.2$ for the lower ($-22.0 \ge b \ge -23.9$, Fig.\ 4b) 
and higher ($-27.1 \ge b \ge 28.1$, Fig.\ 4b) Galactic latitudes, respectively.  
A similar analysis on the summed three middle latitude fields 
m6, a13, and m8 (spanning $-25.1 \ge b \ge -26.2$)
yields $172.3\pm16.5$ stars.  Thus, to within the Poisson errors (and ignoring 
shifts in the portion of the MS luminosity function sampled due to 
distance modulus shifts), there is no  
variation in the density of the feature.  No clear ``center" of highest density
can be discerned within our survey fields.

The above measurements of the surface density of upper MS stars allows an
estimate of the surface brightness of the system via an adopted luminosity 
function.  We actually calculate an upper limit to surface brightness
by adopting the steepest mass functions of Silvestri et al.\ (1998),
namely those with an $x=1$ index where the mass function goes as $m^{-(1+x)}$;
this index assures the most generous contribution of low mass stars for every upper
MS star.  As mentioned above, fitting the 
the $Z=0.0001$ and $Z=0.0004$ Girardi et al.\ (2002) isochrones to our 
ridge line requires a distance modulus of 16.0, so that the 
$21.5 \le M_0 \le 23.0$ range corresponds to the upper
MS magnitudes $5.50 \le M_M \le 7.00$, or $5.33 \le M_V \le 6.77$
after converting to Johnson $V$.
Integrating the  13 Gyr old $Z = 2 \times 10^{-4}$ luminosity function 
(the lowest metallicity computed by Silvestri et al.\ 1998) with 
$x=1$, we find that the upper MS region $5.33 \le M_V \le 6.77$ represents 
12.9\% of the total flux
in this particular luminosity function and we determine that the
$598.1\pm30.6$ detected stars in the feature across our ten fields 
represents a bright limit to the total
surface brightness of the feature of $\Sigma = 32.0\pm0.1$ mag arcsec$^{-2}$.
%
%
If we repeat the calculation
with the Silvestri et al.\ 9 Gyr $Z=0.004$ luminosity function (the
most metal-rich luminosity function they give) we obtain for the 
equivalent $4.83 \le M_V \le 6.27$ MS range (representing 18.3\% of the 
total luminosity function flux) a limit to the surface brightness of
$\Sigma = 32.4\pm0.1$ mag arcsec$^{-2}$. 
%
%
Given that these are lower limits to $\Sigma$, 
derived with the generous $x=1$ mass functions, our detected feature 
is extremely diffuse.  For comparison, the surface brightnesses
of the diffuse, northern parts of the Sgr stream identified by a similar
analysis of MS+MSTO stars by Mart\'inez-Delgado et al.\ (2004) are some
$2.4-4.8$ magnitudes {\it brighter} than even the brighter of the
two $\Sigma$ limits derived above.
If these limits are converted to solar
luminosities at the required distance moduli to match our ridge line, 
we obtain luminosity density limits of $<435.3\pm22.3$ and $<480.8\pm24.6$ L$_{\sun}$ 
deg$^{-2}$ for the $Z=0.0002$ and $Z=0.004$ luminosity functions, respectively.

Accounting for the fact that it also reveals
its presence in the I03 CMDs, this newly 
identified, more distant structure would appear to be at least several times larger
than the roughly $12^{\circ} \times 6^{\circ}$ area we have sampled.  
Recently, by mapping the distribution of 2MASS K and M giants in the Galaxy,
Rocha-Pinto et al.\ (2004, herafter ``R04") have identified an excess, ``cloud-like" 
structure in the same general area of the sky (Triangulum and Andromeda) 
and with the same general properties as the stellar population we
explore here:  Their structure is very diffuse,
lies at several times the distance of the nearby, Monoceros ``ring"  
(which R04 also identify as a separate, distinct structure), and has
a rather smooth density distribution lacking any discernible nucleus  
or core.  Given the overall agreement of general properties,
it is logical to conclude that this structure traced by K and M giants is
the same as that structure traced by the MS+MSTO stars in the present 
contribution.  Some support for this conclusion comes from the radial
velocities of stars that lie within the MS+MSTO feature in M31 CMDs:
In a Keck LRIS spectroscopic survey of the M31 field, 
Reitzel \& Guhathakurta (2002, and in preparation) and Reitzel, Guhathakurta \& Rich (2004)
find four of eight foreground Milky Way ``contaminants" lying along our
upper MS feature (from $21 < M_M < 22.5$) to have radial velocities inconsistent
with the Andromeda Galaxy (M31 stars generally have $v_{hel}<-200$ km s$^{-1}$)
but within 20 km s$^{-1}$ of the mean $v_{hel} = -110 \pm 10$ km s$^{-1}$ (dispersion
18 km s$^{-1}$) that R04 have measured for 
dozens of red giant stars in their ``TriAnd" cloud feature.

If we make the association of the R04 K and M giant structure
with our MS+MSTO structure, then the $50^{\circ} \times 20^{\circ}$ area of the sky
over which R04 detect TriAnd implies
a total system luminosity of a mere
$5 \times 10^5$ L$_{\sun}$, where this is an upper limit within
that area because the luminosity density used is an upper limit.  
This is equivalent to no more than a large 
globular cluster spread over 1000 deg$^2$ of sky.
From these dimensions and mass estimates we must conclude that this 
structure does not resemble any of the known Galactic satellites.
If these stars are tracing the core of a gravitationally self-bound satellite galaxy, 
this ``Triangulum-Andromeda" system must be a rather dark galaxy.

A far more likely scenario, in analogy with previously found
MSTO detections in deep CMDS (e.g., I03; Mart\'inez-Delgado et al.\ 2001, 2004; 
Newberg et al.\ 2002, 2003), is that what we have detected is a piece of
unbound tidal debris in the Galactic halo.  In this case, the feature
is at a distance of $\sim16$-25 kpc, extends more or less uniformly 
across at least a 15$^{\circ}$ span (corresponding to the Fig.\ 1 fields), may
extend to much greater lengths (according to R04), and is 
tenuous compared to tidal debris structures previously identified by their
MS and/or MSTO.
While it is not surprising that many of the most recently found Galactic halo
subcomponents would be clustered at low Galactic latitudes (Willman 
et al.\ 2004), the general location of this new feature of the Milky Way
--- i.e., lying almost directly behind the Newberg-Yanny et al. Monoceros tidal stream 
as identified in Andromeda by I03 --- 
leads one to wonder about possible associations to the latter structure. 
Certainly, as shown in \S3 and R04, the two M31 foreground
features are distinct, but this does not preclude them being
{\it different parts} of the same entity.  As is now well-known
from both N-body models (e.g., Ibata \& Lewis 1998, Johnston et al.\ 1999, 
Helmi et al.\ 2003, Law et al.\ 2004) as well as observations (e.g., Majewski et al.\ 2003)
of tidal debris, extremely long tidal arms from disrupting satellites
can wrap upon themselves, so that lines of sight can intercept multiple 
occurences of the same tidal system at different distances.  Even 
for a satellite disrupting in a nearly circular orbit (an orbit that has been 
suggested for the satellite creating the Newberg-Yanny et al. tidal stream --- 
see, e.g., Crane et al. 2003) the leading/trailing tidal debris arms will 
extend interior/exterior to the parent satellite's actual orbital path, so that
over time a spiral-like configuration of tidal debris can be created.
Further analysis of the chemical and kinematical properties of stars in both the near
and far Andromeda structures, as well as tracing them both to larger
angle on the sky, would help to establish if they are related.
If they are, then the Newberg-Yanny et al.\ stream is extremely long, and, therefore,
relatively old.  We note that while an $\sim20$ kpc distant stream at this position of the
sky is {\it not} predicted by the tidal debris model for the ring 
recently presented by Martin et al.\ (2004), it is fair to 
say that models of the ring are still relatively unconstrained and
immature; pieces of tidal debris as distant as the structure we see here are 
easily accommodated in variants of such models (e.g., see Fig.\ 1
of Helmi et al.\ 2003 and Fig.\ 14 of Martin et al.\ 2004).  


SRM and RJP acknowledge funding by NSF grants AST-0307842 and AST-0307851, NASA/JPL contract
1228235, the David and Lucile Packard Foundation, and The
F.H. Levinson Fund of the Peninsula Community Foundation.
PG acknowledges support from NSF grant AST-0307966 and a
Special Research Grant from UCSC.  PG and DBR thank Linda Pittroff and Drew Phillips for help
  with the LRIS spectral reductions.
We appreciate helpful discussions with Michael Skrutskie.

\begin{figure}
\epsscale{0.69}
\plotone{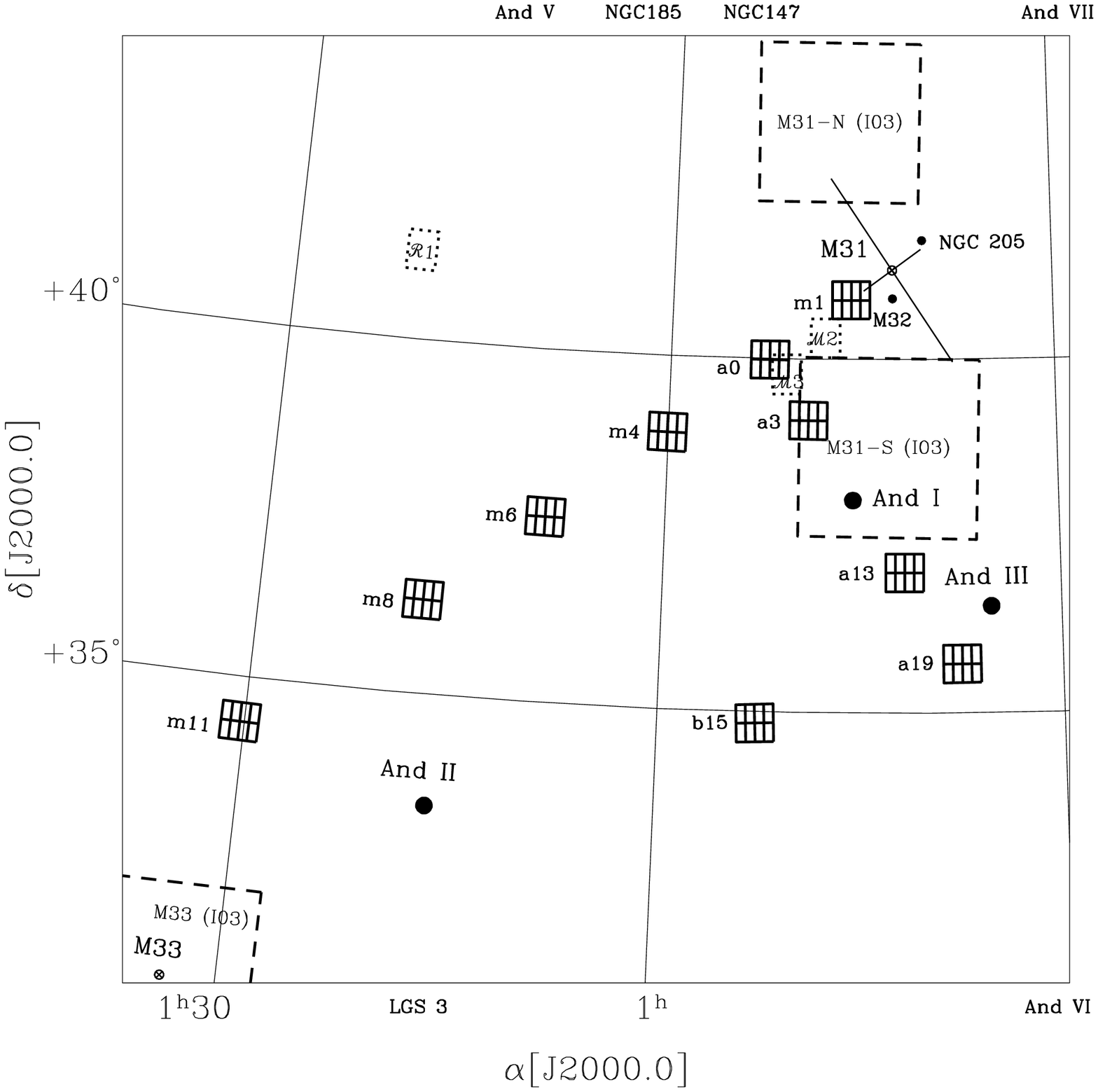}
\caption{Placement of fields studied with the MOSAIC camera in the M31 and M33
region; the fields are shown as the $4\times2$ silhouettes of the individual chips
in the MOSAIC CCD array.  The positions of other Local Group galaxies are also indicated; the features
discussed here are not related to any of these objects.
The fields explored by Durrell et al. (2001) and I03
are also shown as {\it dotted} and {\it dashed} boxes, respectively (the I03
fields are shown only schematically as square fields of the reported
area). }
\end{figure}

\begin{figure}
\epsscale{1.0}
\plotone{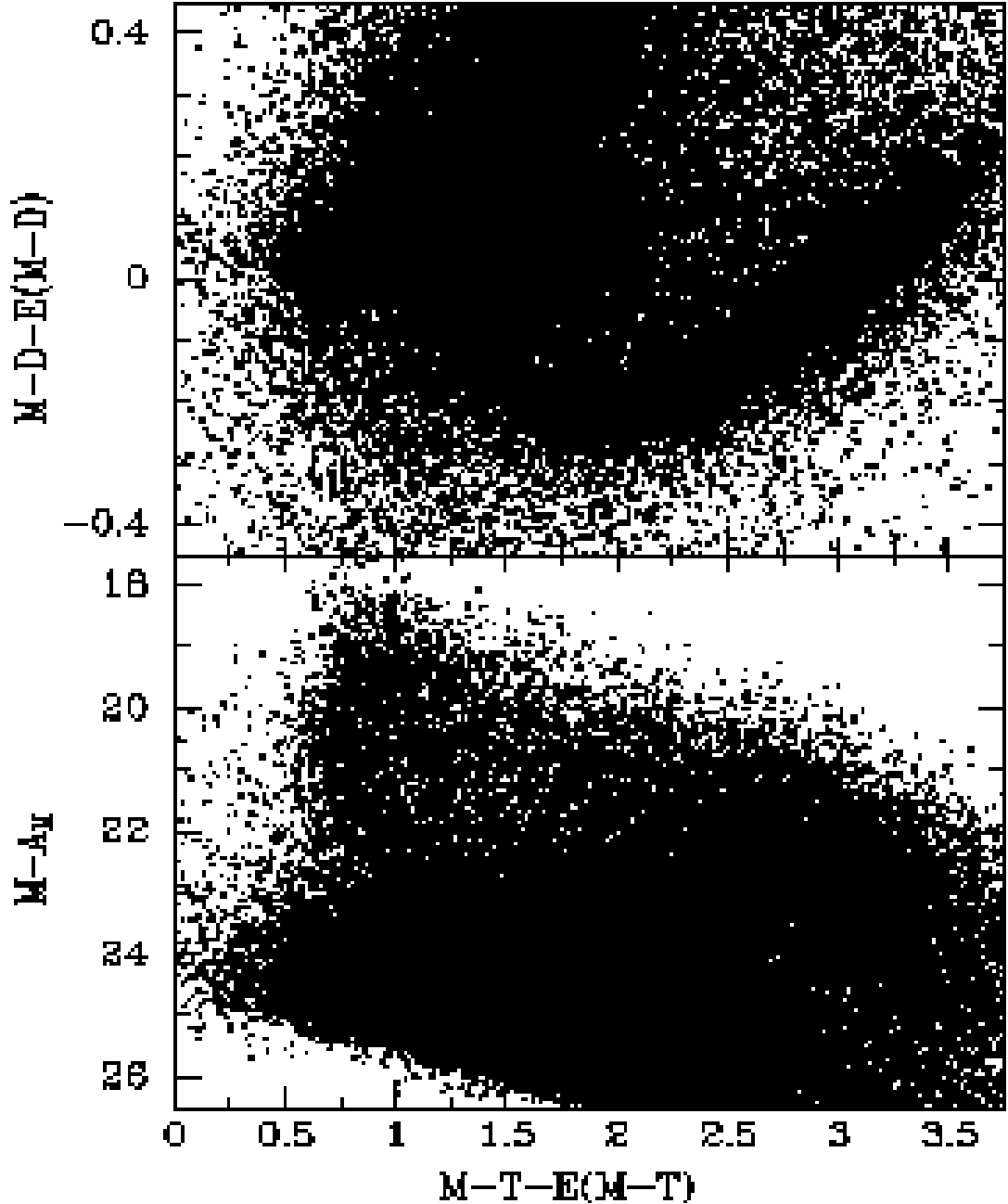}
\caption{{\it Bottom:} Extinction-corrected color-magnitude diagram of more than 8.3 million
sources detected in the ten M31 halo fields.  {\it Top:} The two-color diagram 
(see Majewski et al. 2000), with dereddened colors, for the same stellar sample.}
\end{figure}

\begin{figure}
\epsscale{1.0}
\plotone{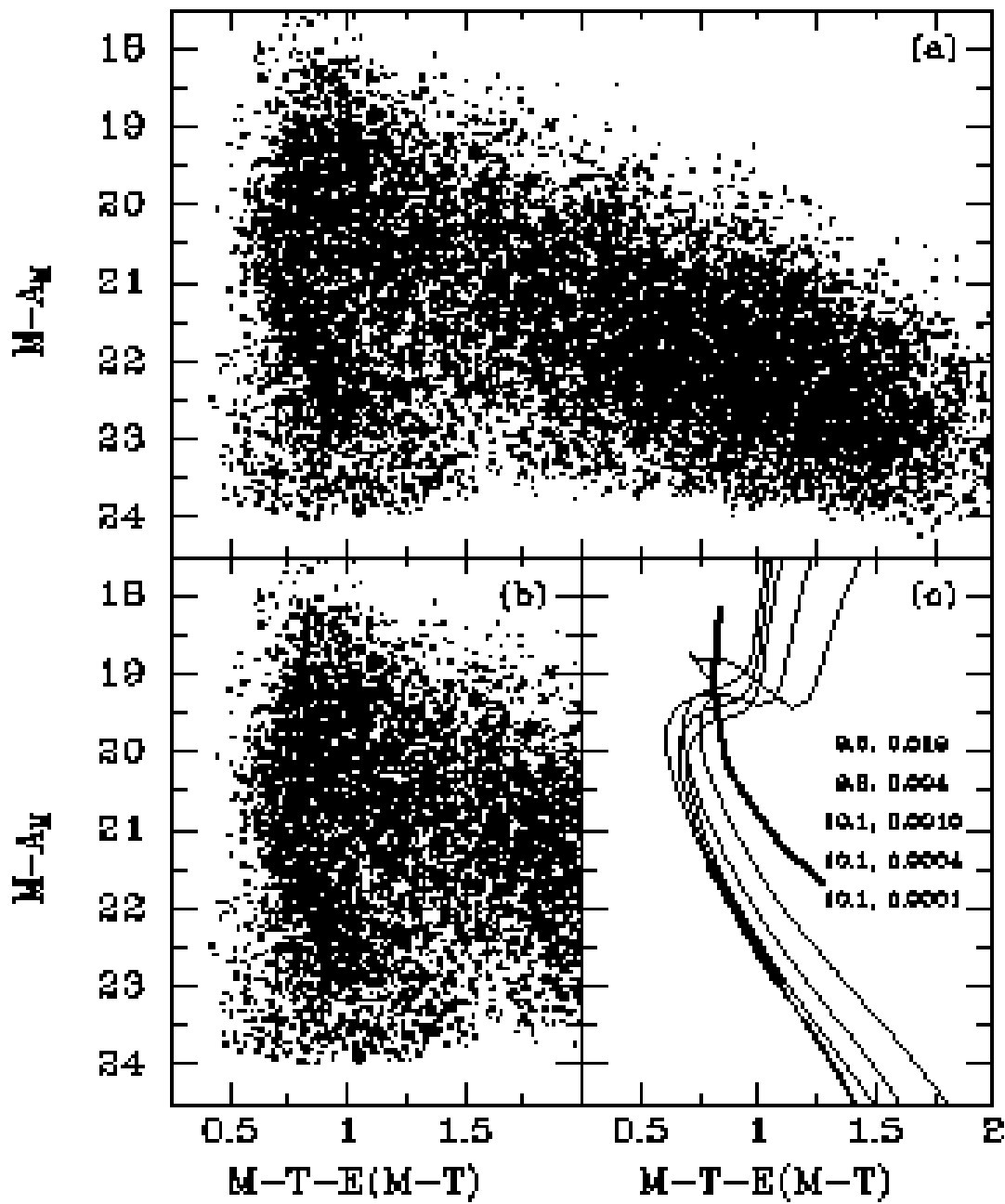}
\caption{Panels (a) and (b) show the CMD
for the ``dwarf-only" sample of 14,000 stars in the M31 fields.
In panels (b) and (c) we show the I03 ridge line for the Monoceros
``ring" (curved upper line) and the MS ridge line used to define the color
distances in Figure 4 (lower straight line).  In panel (c) we include Girardi
et al. (2002) isochrones (thin lines) for five representative combinations of log(age/Gyr) and $Z$,
as shown in the legend (in order of MS brightness). 
}
\end{figure}

\begin{figure}
\epsscale{1.40}
\plotone{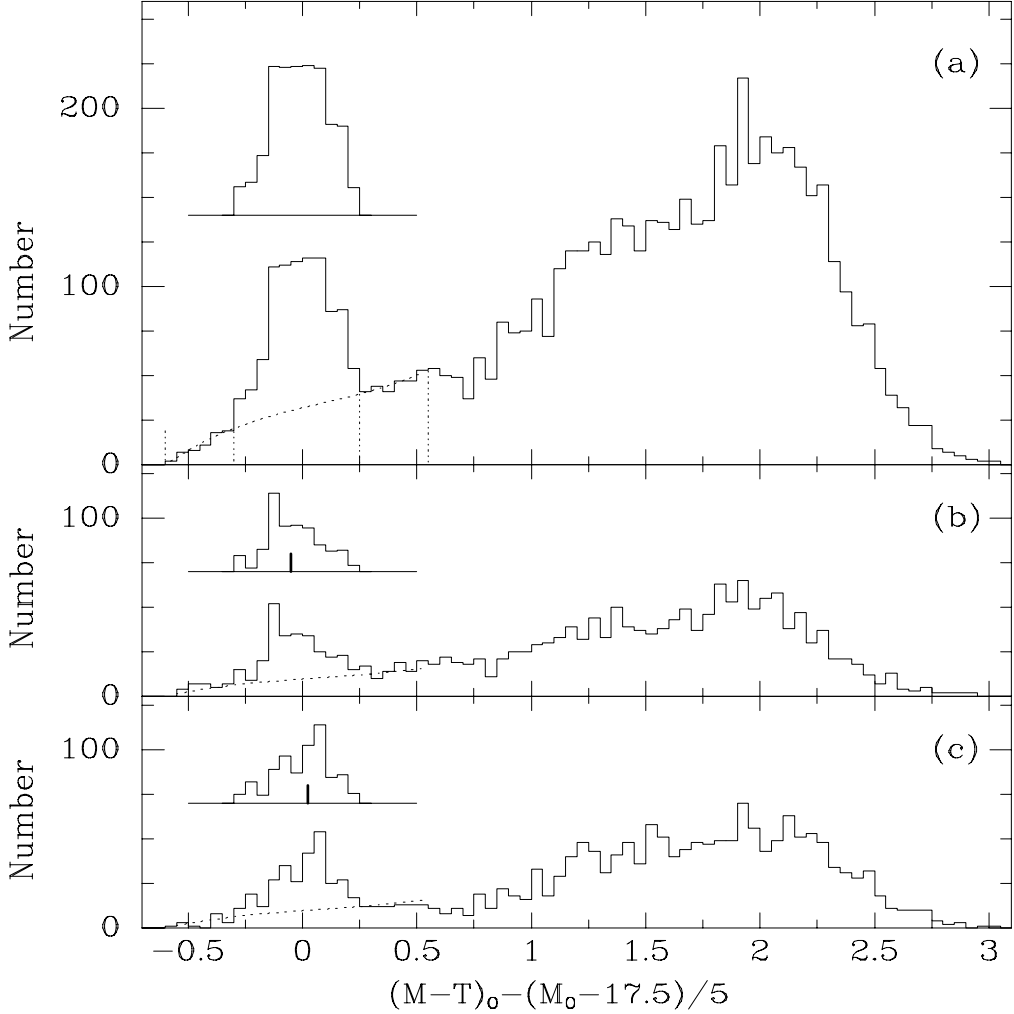}
\caption{Distribution of color differences from the straight ridge line shown in Figure 3
for $21.5 \le M_0 \le 23.0$.  The samples shows are   
(a) all ten fields in our survey (i.e., the stars shown in Figure 3).
(b) the northern fields m1, a0, and m4 combined, 
(c) the southern fields m11, b15 and a19 combined.
The vertical dotted lines delimit the range of color difference used to determine the
background density of stars in the CMD at the position of the MS.  The curving dotted 
line in panel (a) shows the resulting fitted fourth order function; this same
function is shown in panels (b) and (c) after scaling to the area represented.  
The inset to each panel shows
the distribution in the MS peak after subtraction of this background function, with the 
bold line segments in panels (b) and (c) showing the location of the median. 
}
\end{figure}

\begin{figure}
\epsscale{1.0}
\plotone{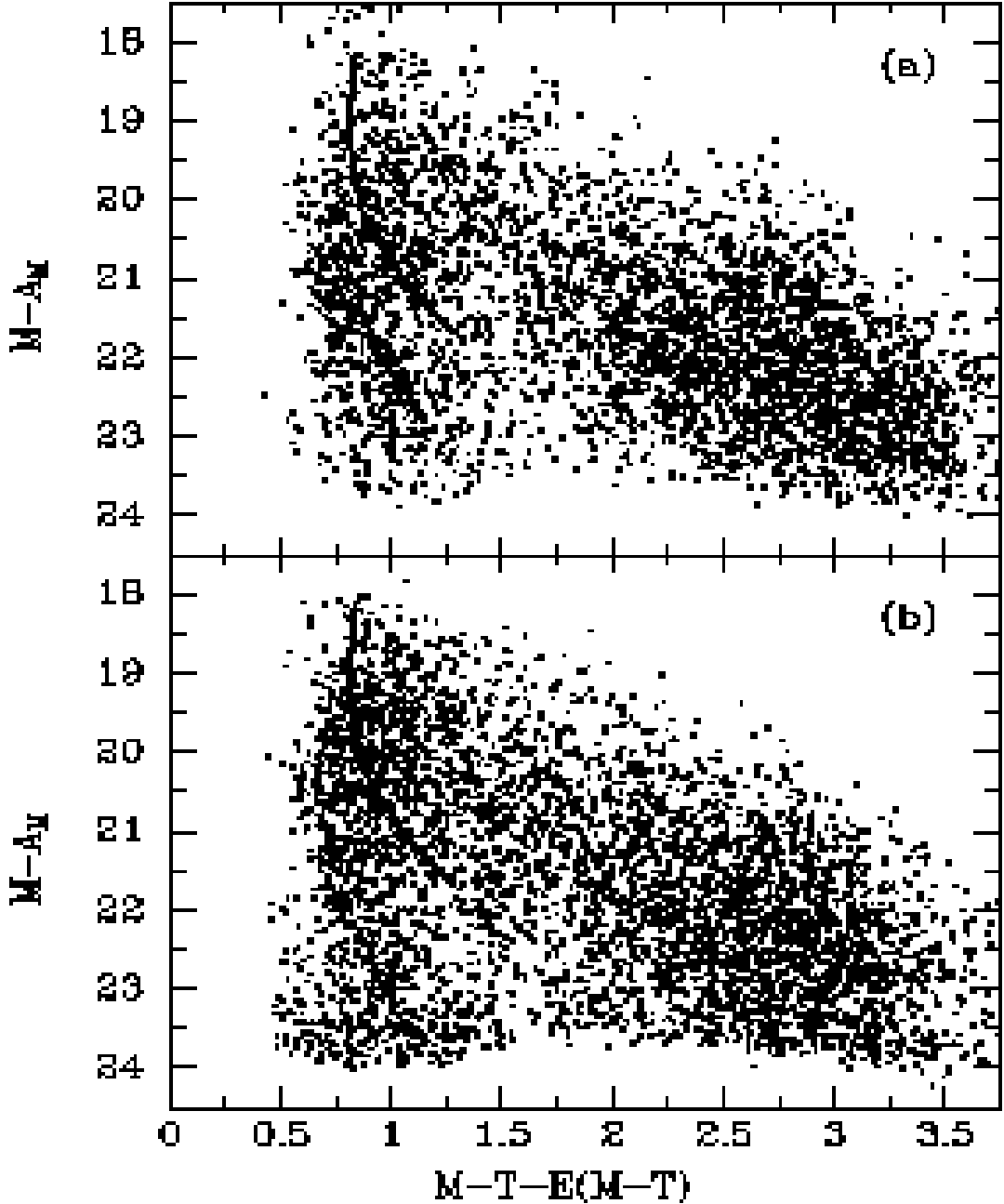}
\caption{$(M-T-E[M-T_2], M-A_M)$ color-magnitude diagram for  ``dwarf-only" samples in ({\it top})
the more northern fields m1,a0 and a4 combined, and ({\it bottom}) the lowest $b$ 
fields m11, b15, a19 combined.  A fiducial reference at $(M-T-E[M-T_2], M-A_M) =
(1.0, 23.0)$ is provided to highlight the difference between the two CMDs.  We include
the I03 ridge line in both panels.
}
\end{figure}

%

\end{document}